\documentstyle[11pt]{article}
\begin{document}
\baselineskip 14pt


\title{The Center-of-Mass of A Few-Body Quantum System\\
            In An Effective Central Field \\
}

\author{Xing Min Wang$^{*}$ and Xiaotong Song$^{\dag}$\\
\\
$^{\dag}$ Institute of Nuclear and High Energy Physics,
Department of Physics,\\
University of Virginia, Charlottesville, VA 22901\\
$^{*}$ Physics Department, Muskingum College, New Concord, OH 43762}

\maketitle

\begin{abstract}
The center-of-mass(CM) of a few-body quantum system with
a central field is discussed. If the particles are in the
designative eigenstates, the CM coordinates of the system can
be well-defined. In the CM bag model as well as in other models
with central fields, the CM-freedom separation rule and
effective nucleon electro-magnetic currents
can be presented without any undetermined parameter.
\end{abstract}

\newpage
It is well known that the static properties of hadrons
can be explained by the nonrelativistic constituent
quark models \cite{ik78}, in which a baryon consists
of three confined valence, constituent, quarks. The
pattern of the hadron spectroscopy fitted nicely in
the symmetry group $SU(3)_{flavor}\otimes SU(2)_{spin}$,
together with the $O(3)$-nonrelativistic oscillator
spatial wave functions. In this case, the center-of-mass
(CM) motion can well be separated from the internal
relative motion. However, the motion of the light quarks,
e.g. up and down quarks in the nucleon is highly
relativistic because the kinetic energy of a quark
is almost the same as its constituent mass. Hadron models,
which can account for relativistic motion of the quarks,
lost an advantage of the nonrelativistic quark model due
to the intrinsic non-separability of the center-of-mass
motion for a relativistic 3-body system. For the MIT bag
model \cite{MIT1}, the CM is at rest and the bag is static,
hence the CM degree of freedom has been completely
disregarded. This does not matter for describing the mass
spectroscopy of the hadrons, but does matter for the hadron
decay and scattering processes where the recoil effect, hence
the CM motion cannot be neglected. One of the consequences
of neglecting the CM degree of freedom is that the
translational invariance and 4-momentum conservation
are lost.\cite{MIT2}

To recover the 4-momentum conservation, Barnhill III
has made a proposal \cite{Barn} that the wave-function of
a 3-quark system has a collective plane-wave factor:

\begin{equation}
\phi_P(y) =\frac{1}{ (2\pi)^{3/2}}e^{-iP \cdot y}
\label{eq:phipy}
\end{equation}
where $y^\mu=(T,{\bf y})$ is the center-of-mass(CM) coordinates
and $P^\mu$ is the total 4-momentum of the system. In \cite{CM-0}
we have modified and generalized this assumption and developed
a formalism in the calculation of hadron structure functions
and electromagnetic form-factors \cite{CM-01,CM-1}. We also
provided some field-theory
basis of this idea and proposed the so-called CM bag model
\cite{CM-2}. Some Feynman rules and their applications, which
includes a possible explanation of the nuclear EMC effect,
were presented \cite{CM-3,CM-4}. Recently, considering the
symmetry breaking effects coming from the spin dependent
quark-quark interactions, the magnetic moments of baryons,
elastic form factors and deep inelastic structure functions
of the nucleon are calculated \cite{CM-5}. We note that
there is another approach in the literature to
avoid the disadvantage of the original MIT bag model, e.g.
see Ref. \cite{bm}. In this approach, the Peirels-Yoccoz
projection \cite{py} is used to obtain an eigenstate of
zero momentum.

In ref \cite{CM-2}, we gave a brief discussion of how we
introduce the plane wave function for the CM degrees of
freedom. The CM freedom separation rule is:
( see eq. (20b) in ref.\cite{CM-2}):

\begin{equation}
\Psi(x_1,x_2,x_3) =\sum_n \int d^3P \lambda^{3/2} a_P
\frac{e^{-iP \cdot y}}{ (2\pi)^{3/2}} b_n\,q_{np}(\xi_1, \xi_2, \xi_3)
\label{eq:sep_rule0}
\end{equation}
where $\lambda $ is an undetermined factor of dimension $ [\lambda ] =L $
(length).

In this letter, we will discuss in more detail about our basic
assumptions on the 4-potential $V_\mu$, the restrictions on the
states of particles, the definition and the {\em revised} separation
rule of the CM-coordinates $y^\mu$ for a non-relativistic or
relativistic N-particle quantum system. The parameter $\lambda$
does not appear in our revised separation rule, nor in revised
effective electromagnetic current of the nucleon. It appears
only when we use the free-quark approximation for outgoing quarks
in deep inelastic collisions. Our method can be applied to other
hadron or nuclear models with central fields.

\vspace{0.4cm}
    {\it 1.  The 4-potential and the 4-momenta:} Our main assumption
is that there are $N$ particles moving in an effective 4-potential:
$ V^{\mu}_a\equiv V^\mu(x_a) $, which is reduced to a stable central
field in a  CM-rest reference frame(CMRF), i.e., where the total
3-momentum of the particle system vanishes. The center of the field
is always located at the CM  position ${\bf y}$. Therefore, in a CMRF,
we can write:

 \begin{eqnarray}
V^{\mu}_a &\equiv & V^\mu(r_a) = (V(r_a), {\bf 0} ) \label{eq:Gmu1}\\
 r_a &=& \left|x_a - y \right|_{t_a =T} = \left|{\bf x}_a -
{\bf y}\right|
\equiv\left| {\bf \xi}_a \right|
\label{eq:Gmu2}
\end{eqnarray}
Here we put  $ t_a= T $, because $r_a$ are proper lengths and the
events $x_a$ and $ y$ have to always  be simultaneously measured
in a CMRF. This implies that,  whatever the definition of the CM
4-vector $y^\mu = (T, {\bf y})$ is, $T = T(t_1,..., t_N)$ has to
contain the following special solution:

\begin{equation}
t_1 = t_2  = \cdots = t_N = T  \hspace{0.5cm} \mbox{in a CMRF}
\label{eq:ta=T}
\end{equation}
If  it is  a non-relativistic N-body system, this of course must
be true. Moreover, when we say there is a central field centered
at the CM position, we mean that the CM position is fixed, i.e.,
${\bf y }= \langle {\bf y } \rangle $. Since $ y^\mu$ is 4-vector,
we have the second requirement on CM:

\begin{equation}
 y^\mu  = \langle y^\mu \rangle \equiv Y^\mu
 \label{eq:y=Y}
\end{equation}
To be consistent with our assumption on $V^\mu$,  we have to put
some restrictions on the states of particles.  For example, to make
the effective interaction field stable in a CMRF, we should only
consider the stationary states or energy eigenstates.  Moreover,
because the effective field is isotropic, it is reasonable to
restrict to states with zero expectation value of $x_a - y$, i.e.,
to eigenstates of parity.  We call such states {\em designative states}.

In  quantum mechanics, when we use the coordinate representation
(or, the Schr\"{o}dinger picture),  the 4-momentum become operator,
$ \hat{p}_\mu \equiv i\partial_\mu$, and we have  the commutators:
 $[ x_a^\mu, \hat{p}_{b\,\nu}] = - i \delta^\mu_\nu\, \delta_{ab}$.
In a given state $  \psi (x) $, we have the expectation values of the
momentum:

\[ \bar{p}_\mu = \int d^3 x  \psi ^{\ast}(x)  i\partial_\mu  \psi(x) \]
The eigenvalue of total 4-momentum  $\hat{P}_\nu$ of the system is
denoted by:

\begin{equation}
P^\mu = ( \sum_{a} \epsilon_a, \sum_{a} {\bf p}_a) = (E, {\bf P})
\label{eq:Pmu}
\end{equation}
$\hat{P}_\nu$ and $y^\mu$ must be mutually canonically
conjugate, or:

\begin{equation}
 \left[ y^\mu, \hat{P}_\nu \right] = - i \delta^\mu_\nu
\label{eq:[y,P]}
\end{equation}
In the Schr\"{o}dinger picture, $ \hat{P}^0 =
\sum_{a} \hat{p}^0_a = \sum_{a}  i
\partial_{t_a}$,  we have  the third  requirement on the definition
of CM time $T$, i.e.,  for any function  $F(T)$,

\begin{equation}
  \partial_T  F(T) = \sum_{a}   \partial_{t_a} F(T(t_1, \ldots, t_N))
\label{eq:TF(T)}
\end{equation}
Now let $ F(T) = T$, then eq.(\ref{eq:TF(T)}) shows that $T$ must be
linear in $t_a$, with $  y^0 = T = \sum_{a}\eta_a t_a$
where   $ \sum_{a}\eta_a = 1 $ .
Now that $y^\mu$ is a 4-vector, we have
\begin{equation}
 y^\mu =  \sum_{a}  \eta_a x_a^\mu\ , \hspace{0.6cm} \sum_{a} \eta_a = 1
\label{eq:yeta}
\end{equation}
where $\eta_a$  have to be  Lorentz scalars. Eq.(\ref{eq:yeta}) is also
consistent with eq.(\ref{eq:ta=T}).

One might want to define $ \eta_a = m_a/\sum_{a} m_a$, where $m_a$
is the rest mass of the $a$-th particle. Such a definition has two
disadvantages here. First, it does not work for massless particles,
like the quarks in the MIT bag model. Second, in the case of a central
field with its center located at the CM of the system, it is impossible
to solve for the wave functions exactly. In next two sections, we will
see that it is more naturally to define $ \eta_a=\epsilon_a/\sum
\epsilon_a$, where $\epsilon_a$ is the energy eigenvalue of the $a$-th
particle.

In a CMRF, $P^\mu$ and $P\cdot y$,  the most important scalar to
describe the motion of the CM due to eq.(\ref{eq:phipy}), are reduced
to:

\begin{equation}
 P^\mu = ( \sum_{a} \epsilon^{(0)}_a, {\bf 0})
=(\sum_{a} \omega_a, {\bf 0})=  (M, {\bf 0})
\label{eq:P0}
\end{equation}
\begin{equation}
 P \cdot y = ( \sum_{a} \omega_a)T =  M \, T
\label{eq:P.y1}
\end{equation}
Here $M$ should be taken as the rest mass of the system. Now let CMVF
be a new frame, moving in the $ -{\bf x}_{\|}$ direction at a velocity,
$ {\bf v} = ( -v, {\bf 0}_{\bot} )$  with respect to the CMRF, then we
have the Lorentz transformation for $ \tilde{x} =\Lambda x $, with

\begin{equation}
\Lambda = \left( \begin {array}{rrr}
      \gamma & v \gamma  & 0  \\   v \gamma & \gamma  & 0  \\
              0 & 0 & I_\bot \end{array}  \right) ,
           \hspace{0.5cm}
    \Lambda^{-1} = \left( \begin {array}{rrr}
   \gamma & -  v \gamma  & 0  \\  - v \gamma & \gamma  & 0  \\
         0& 0 & I_\bot    \end{array}  \right) ,
                 \hspace{0.5cm} \gamma = \frac{1}{\sqrt{1 - v^2}}
\label{eq:Lambda}
\end{equation}
In the CMVF,  we find
$P^\mu  \rightarrow   \tilde{P}^\mu = ( \gamma M ,\gamma M{\bf v })$,
and

\begin{equation}
 P \cdot y =  \tilde{P} \cdot \tilde{y} = \gamma M\,\tilde{T} -\gamma M
\,  {\bf v }\cdot \tilde{\bf y}
\label{eq:P.y2}
\end{equation}

\vspace{0.4cm}
{\it  2.  The CM 4-vector of a Classical Steady Few-body System:}
Before we go to quantum system, let us consider classical relativistic
system first. According to our restrictions on the motion of particles,
the closest classical cases are so-called steady or rigid-body solutions
\cite{Steady}, when particles moving periodically around the center of
mass in fixed orbits with fixed speed and energy in a CMRF:

\begin{equation}
\omega_a = \mbox{const.}\ ,  \hspace{0.8cm} {\bf x}_a (\tau + S) = {\bf
x}_a(\tau)
\label{eq:period}
\end{equation}
Similar to the expectation values of momentum in quantum mechanics, here
we consider, instead of the instantaneous momentum, the time-averaged
4-momentum of each particle,

\begin{equation}
\bar{p}_a^\mu = \frac{1}{S} \int_{0}^{S}  p_a^\mu (\tau)\;d\tau
\label{eq:pbar1}
\end{equation}
Note that $\bar{p}_a^\mu $ is also a 4-vector by definition, and in
a CMRF it reduced to:

\begin{equation}
\bar{p}_a^\mu = (\omega_a, {\bf 0} )
\label{eq:pbar2}
\end{equation}
We see that the rest energy $\omega_a = \sqrt{\bar {p}^2}$ has to be
considered as a Lorentz-invariant constant. The time-averaged total
momentum of the system in a CMRF is the same as the instantaneous
total momentum:

\begin{equation}
\bar{P}^\mu = (\sum \omega_a, {\bf 0} ) = (M, {\bf 0}) = P^\mu
\label{eq:pbar3}
\end{equation}
Note, as a 4-vector equation, $\bar{P}^\mu = P^\mu$ is true in any
CMVF and $M$ is also a Lorentz invariant constant. This enable us
to define the CM 4-vector by:

\begin{equation}
y^\mu = \frac{\sum \omega_a x^\mu_a}{\sum \omega_a}
             = \frac{1}{M}\, \sum_{a}\omega_a x^\mu_a =
\sum_{a} \eta_a x_a^\mu
\label{eq:CMD2c}
\end{equation}
By our definition, $y^\mu$ is a Lorentz 4-vector. We see that
$ \bar {\bf p}_a ={\bf 0} $ is crucial to our definition.
It is easy to verify  that  this is consistent with eqs.(\ref{eq:ta=T}),
(\ref{eq:yeta}), and the Poisson brackets between $y^\mu$ and $P_\nu$
are satisfied.  Besides, in a CMRF, $ {\bf y } = \bar{\bf y}
={\bf Y} $,  because we have chosen steady periodic solutions.
Also, $ T = \bar{T}$ because $T$ and $t_a$'s are linear in $\tau$.
Therefore, the requirement of eq.(\ref{eq:y=Y}) is valid.

In a CMRF, we have $ P \cdot y = M \, T = \sum\omega_a t_a $, as given
by eq.(\ref{eq:P.y1}). In a CMVF, $P \cdot y$ as a scalar is given by
eq.(\ref{eq:P.y2}). But we can also obtain it by using the inverse
Lorentz transformation $t_a=(\Lambda^{-1})^0_{\mu}{\tilde x}^{\mu}$,
taking into account that $\omega_a $ are Lorentz-invariants and that
by definition:

\[ \sum_{a} \omega_a \tilde{x}_{a \,\|} = (\sum_{a} \omega_a) \tilde{y}_{\|} \]
Because $\epsilon_a=\gamma\omega_a$ for each $a$ in a CMVF, our
definition (\ref{eq:CMD2c}) is equivalent to:
\begin{equation}
y^\mu = \frac{\sum \epsilon_a x^\mu_a}{\sum \epsilon_a}
\label{eq:CMD3}
\end{equation}
which reflects the fact that $y^\mu$ is a 4-vector and we have used this
notation in ref. \cite{CM-2}.

\vspace{0.4cm}
{\it 3.  The CM 4-vector of a Non-relativistic  Quantum System}: Now
let us turn to non-relativistic quantum systems. We will see that
many features of the CM degrees of freedom can be revealed in such
cases and they may have important applications in atomic and nuclear
physics. We will still use our 4-vector notation, though Lorentz
covariance is not a requirement.

Under our assumption on the 4-potential, in a CMRF each particle has
the following Schr\"{o}dinger equation:

\[  i \partial_{t_a} \psi(x_a) = H(x_a) \psi(x_a)
      = [ m_a + \frac{1}{2m_a}{\bf  \hat{p}}_a^{\;2} +V(r_a)   ]\psi(x_a) \]
where we have temporally used the  notation $t_a$,  which has to
satisfy eq.(\ref{eq:ta=T}) in the non-relativistic limit. We also
have assumed that ${\bf y}_a={\bf 0}$ in the CMRF. We consider only
$designative$ states, or bounded energy eigenstate with definite
parity. They can be written as $\psi_n(x_a)=\exp(-i\omega_a t_a)
q_n( \xi_a)$ and normalized as:

\[ \int  d^3 x_a  \psi_n^\ast ( x_a)   \psi_n(x_a)  = 1 \]
The expectation values of momentum and coordinates are:

\[  \langle \hat{p}_a^\mu \rangle =  \bar{ p}_a ^\mu = \int  d^3 x_a
\psi_n^\ast ( x_a)  i \frac{ \partial}{\partial x_a^\mu} \psi_n(x_a)  = (
\omega_a, {\bf 0})  \]
\[  \langle x_a^\mu \rangle  = \bar{ x}_a ^\mu = ( t_a, {\bf 0}) \]

Now we can define the CM 4-vector as in eq.(\ref{eq:CMD2c}). Here
again, the definition of $T$ is consistent with eqs.(\ref{eq:ta=T}),
(\ref{eq:TF(T)}), which are, of course, true in non-relativistic
theories. We also have $y^\mu=\langle y^\mu \rangle=Y^\mu$  as in
eq. (\ref{eq:y=Y}).  Moreover, from our definition of $y^\mu$, one
can easily find that $y^\mu$ and $P_\nu$ are canonically conjugate
and for any function $F(y)$,  we have:

\begin{equation}
 \sum_{a} \frac{\partial}{\partial x_a^\mu}F[y(x_1,...,x_N)]
     = \frac{\partial}{\partial y^\mu}F(y)
\label{eq:patialFy}
\end{equation}

In a stable central field, we can find states $ \psi_{\check{n}}(x)$
of definite energy and angular momentum $ {\bf L }^2$ , $ L_z $ and
$S_z$, with corresponding quantum numbers $ \check{n}\equiv (n, l, m, m_s)$.
In these states, we have the required expectation values of
momentum and coordinates to use our CM definition. The product
of N single particle wave functions leads to:

\[\prod_a \psi_{\check{n}_a}(x_a) \equiv
     e^{ -i \sum_{a} \omega_a t_a} \prod_{a} q_{\check{n}_a}( \xi_a)
    = e^ { -i M T } \prod_{a} q_{\check{n}_a} ( \xi_a)  \equiv
 e^ { -i M T } q_{\check{n}}(\xi_1, \ldots  \xi_N) \]
The spatial part of the plane wave function
$ \exp( i {\bf P} \cdot {\bf y} )$ disappears because
${\bf P} = {\bf 0}$.
Now let us check if $dY^i/dT$ represents the motion of the CM in
a CMVR, where $ x^i_a = \xi^i_a + v^iT $ and there is no change
in $H(x^i_a) = H(\xi^i_a)$. We find:

\[  \frac{dY^i}{dT} =\sum_{a} \eta_a \frac{d \langle  x_a^i \rangle }{dT}
=\sum_{a} \eta_a \langle  \frac{\partial x_a^i }{\partial T}
+ i[ H, x_a^i]\rangle=\sum_{a} \eta_a v^i = v^i  \]
So the system does have a total 3-momentum $ P^i = M dY^i/dT = Mv^i $
and a total energy $ E = M + Mv^2 /2$ as expected.  To recover
$ \exp(-iP \cdot Y) $ in a CMVF, we first apply Lorentz transformation
to $ M\,T = P \cdot Y$ and then take the non-relativistic limit
$ v\ll 1 $, to obtain the wave function in a CMVR (this could  be
called a {\em semi-relativistic} treatment):

\begin{eqnarray}
\Phi_{\check{n}P}(\xi_1,...,\xi_N;Y) & \equiv
& \frac{1}{(2\pi)^{3/2}} e^{-iP \cdot Y} q_{\check{n}}(\xi_1, \ldots \xi_N)
\label{eq:Phi_nP}  \\
\int d^3 Y d^{3N}\xi \, \Phi^\ast_{\check{n} P}\Phi_{\check{n}' P'} &=&
\delta_{\check{n}, \check{n}'} \delta^3({\bf P}-{\bf P}')
\label{eq:PhiP-P'}
\end{eqnarray}
where $ P^0 = E=\gamma M \approx M + Mv^2/2+ \cdots $ , $ P^i = \gamma
M v ^i \approx M v^i + \cdots $,  as we expected, and  $d^{3N}\xi = d^3\xi_1
\cdots d^3\xi_N$ .

     How can we  consider $Y^i$ as independent variables? It is
important to note  that,  because $y^\mu = Y^\mu$  in the CMRF,
we have the following three functional restrictions on the states:

\begin{equation}
y^\mu \equiv \sum_{a} \omega_a x_a^i = \sum_{a} \omega_a \langle  x_a^i
\rangle \equiv Y^\mu
\label{eq:y=Yf}
\end{equation}
These are very strong restrictions, implied  by our assumption: the
center of mass is the center of the potential. We see that the
designative system states are not simply the product of any
individual particle states. For example:
\[ [c_a \psi_a(x_1) + c_b \psi_b(x_1)]\psi_{n_2}(x_2) \cdots
    \psi_{n_N}(x_N) \]
is a solution of the N-particle Schr\"{o}dinger equation, but not
an energy eigenstate  of the system  if  $ \omega_a \neq \omega_b $.
Thus, in choosing our designative states, we have already reduced
the degrees of freedom.

Now we want to expand any $ \Psi(x_1, \ldots, x_N) = \langle x_1,
\ldots, x_N|\Psi \rangle$  of the system by using our orthogonal
function set (\ref{eq:Phi_nP}), while keeping $ \langle \Psi |
\Psi \rangle = 1 $. Since $ {\bf P } = M {\bf v}$ and

\[  \langle {\bf P } | {\bf P }' \rangle = \delta^3( {\bf P }
- {\bf P }') =\delta^3( {\bf v } - {\bf v }')/M^3 \]
we introduce:

\begin{eqnarray}
 |{\bf v}\rangle &=& (M)^{3/2} | {\bf P } \rangle \\
\phi_v( { Y}) &=& (M)^{3/2} \phi_P({ Y}) = \frac{(M)^{3/2}}{(2\pi)^{3/2}}
 e^{ - i P \cdot Y}
\label{eq:phi_v}
\end{eqnarray}
with $ P^\mu = (M, M {\bf v})$ in non-relativistic limit. Thus we
have the following {\em semi-relativistic CM-freedom separation rule}:

\begin{eqnarray}
\Psi(x_1,.., x_N) & \equiv &  \Psi(\xi_1,.., \xi_N;Y)
       =\sum _{\check{n}}  \int d^3 v \, a_v \phi_v(Y)  b_{\check{n}}
q_{\check{n}}(\xi_1,..,  \xi_N)     \\
    & =& \sum _{\check{n}}  \int d^3 P \, a_P \phi_P(Y)
          b_{\check{n}} q_{\check{n}}(\xi_1, \ldots,  \xi_N) \\
   & \equiv & \sum _{\check{n}}  \int d^3 P \,
      a_P b_{\check{n}} \Psi_{\check{n}P}(\xi_1, \ldots; Y)
\label{eq:sep_rule1}
\end{eqnarray}
where we have used the facts that $  d^3 v =  d^3 P /M^3$ and $ a_v =
M^{3/2}a_P $ . This separation rule has several advantages:  it has no
undetermined parameter $\lambda$, as in eq.(\ref{eq:sep_rule0}); it
has the right normalization: $ \int d^3Y\, d^{3N}\xi \Psi^\ast \Psi = 1$,
if  $ \int d^3P |a_P|^2 = 1$;  and it has right dimension $ [ |\Psi|^2 ]
= L^{-3N-3} $ to fit the requirement from the field theory later.
We should keep in mind the dimensional relation:

\begin{equation}
 [ \psi(x_1) \cdots \psi(x_N)  ]= L^{-3N/2}= L^{3/2} [ \Psi(x_1,\ldots,x_N)]
\label{eq:dimension}
\end{equation}

In single-particle problems, when the perturbation Hamiltonian is $H'(x)$,
initial state is $ | \psi_n \rangle $ and final state is $| \psi_{n'}
\rangle$,  the S-matrix element is $ \langle \psi_{n'}|H'| \psi_{n}
\rangle $. This  can be readily extended to our confined N-body system:
\begin{equation}
\langle \Psi_{n'P'}  |H' | \Psi_{nP} \rangle
\label{eq:H'PP'}
\end{equation}
For example, let us look at a simple example:  a polarized (in x-direction)
$\gamma$-ray  traveling in ${\bf k}$-direction, with electric field
$E_x=E_0Re(e^{-i k\cdot x})=E_0(e^{-ik\cdot x}+e^{ik\cdot x})/2$, is
interacting with the $N_{p}$ protons in a nucleus of $N$ nucleons at
initial state $ \psi_{n_i}(x_i) = \exp( -i \omega_i t_i)q_{n_i}(\xi_i)$
in an isotropic harmonic oscillator potential $V(r)$ (not necessarily
being relativistic solutions). The perturbation then is:
\begin{eqnarray*}
 H' &=&\sum_{i=1}^{N_p} -e
({\bf x}_i - {\bf y}) \cdot {\bf E} \,Re(e^{-i k \cdot x_i})  \\
& =& \sum_{i=1}^{N_p}- e E_0 \, \xi_{ix}\,Re( e^{-i k \cdot y}
e^{i {\bf k} \cdot {\bf \xi}_i})
\end{eqnarray*}
where we have used $ {\bf x}_i ={\bf y}_i +{\bf \xi}_i$ and $t_i=T$.
The S-matrix element is:

\[  \langle  \Psi_{n'P'}| H' | \Psi_{nP} \rangle =\sum_{i=1}^{N_p} \int d^4 Y
d^3{\vec\xi}_i \frac{e^{- i(P -P')\cdot Y}}{(2 \pi)^3} q_{n'_i}^\ast ({\xi}_i)
H'(x_1) q_{n_i}({\xi}_i) \prod_{j \neq i}^{N} q^\ast _{n_j} q_{n_j} \]
which leads to:

\[  \frac{1}{2}(2\pi)e E_0 \delta^4(P+k -P')
     \sum_{i=1}^{N_p} \int d^3{\vec\xi}_i
e^{i {\bf k} \cdot {\vec\xi}_{i}} q_{n'_i}^\ast (\xi_i)
{\xi}_{ix} q_{n_i}({\xi}_i) \]
(there is another term with a factor $ \delta^4(P-k -P')$, which is
always zero). From this equation we can easily find the recoil of
the nucleus and the allowed change of states. For, example, let $P^\mu=
(M, {\bf 0})$,  then $\epsilon_{n'_i} - \epsilon_{n_i} = k - k^2/(2M)$,
which is not that when the recoil is neglected. One can also see that
there is no elastic scattering, when $ n'_i=n_i$ and $P'^2=P^2=M^2$,
as is well known in Compton effect.

\vspace{0.4cm}
{\it 4.  The CM 4-vector of  a relativistic Quantum System}:  Now we
are ready to discuss  the CM 4-vector of a relativistic system with N
spin-1/2 fermions, confined in a central field.  We want to use the
expectation values of $p^\mu$ and $x^\mu$, thus $ | \psi(x)|^2 $ should
be still interpreted as the probability distribution or particle density
in the space. This is true if we only concentrate on particles (quarks
or nucleons) and avoid the particle-antiparticle creation or annihilation.
We use the Schr\"{o}dinger picture where the states of particle are
spinor functions of coordinates $x_a^\mu$, satisfying  the following
Dirac equations:

\begin{equation}
\gamma^\mu ( i \partial_{a\, \mu}  -  G_\mu (x_a) ) \psi (x_a)
       = m_a \psi(x_a)
\label{eq:Dirac}
\end{equation}

In a CMRF,  the 4-potential is reduced to a time-independent central
field, the designative states can be chosen as $
\psi_{n_a}(x_a)$  with definite energy $ E_{n_a} = \omega_a $, satisfying:

\begin{eqnarray}
i \partial_{a\,0}\psi_{n_a}(x_a) &=& \omega_a\, \psi_{n_a}(x_a) \\
\psi_{n_a}(x_a) &=& e^{-i \omega_at_a}\, q_{n_a}({\xi}_a)
\label{eq:psix}
\end{eqnarray}
The wave functions can be normalized as:

\[ \int d^3 x_a \psi^\dagger_{n_a}(x_a) \psi_{n'_a}(x_a)
= \delta_{n_a \, n'_a} \]
When $n_a = n'_a$,  it is the conserved total ``charge".  Now we have  the
expectation values of  $p^\mu_a$  and $x^\mu_a$  in the designative states:

\begin{equation}
 \bar{p}_a^\mu =  \int d^3x_a   \psi^\dagger_{n_a} i \partial ^\mu_a
\psi_{n_a} =  \langle \hat{p} ^\mu_a \rangle =( \epsilon_a, {\bf 0})
\label{eq:pbarqm}
\end{equation}
\begin{equation}
\bar{x}_a^\mu =  \int d^3x_a \psi ^\dagger_{n_a} i x^\mu
\psi_{n_a}= \langle x ^\mu_a \rangle = ( t_a, {\bf 0})
\label{eq:qbarqm}
\end{equation}
again, these are 4-vectors by definition. The expectation value of
total 4-momentum is $(M,{\bf 0 })$. These relations enable us to use
eq.(\ref{eq:CMD2c}) as the definition of  $y^\mu $  and obtain the
product of designative states of the N-particle system:

\begin{eqnarray}
\prod_{a}^{N} \psi_{n_a}(x_a) &=& e^{-i \sum \omega_a t_a}\,
\prod_{a}^{N}q_{n_a}({\xi}_a)  \\
&=& e^{-i M\, T}\prod_{a}^{N}q_{n_a}({\xi}_a)
=  e^{-i M\, T}q_n({\xi}_1, \ldots, {\xi}_N)
\label{eq:3psi1}
\end{eqnarray}
Transformed to a CMVF,  $ M \,T$ becomes $P \cdot Y $ through Lorentz
transformation, as we did before,  and the product of wave functions
takes the form (with a normalization coefficient):

\begin{eqnarray}
\psi_{nP}(x_1, \ldots, x_N) &=&  \frac{1}{(2\pi)^{3/2 }} e^{-i P \cdot  Y}
  \prod_{a}^{N}S(\Lambda_v) q_{n_a}({\xi}'_a) \\
 \equiv  \phi_P(Y)  \prod_{a}^{N}q_{n_a v}({\xi}_a) &\equiv &
\phi_P(Y) q_{nv}({\xi}_1, \ldots, {\xi}_N)
\label{eq:3psi2}
\end{eqnarray}
where $ \xi' = \Lambda^{-1}\xi = (\gamma \xi_\|, \xi_\bot)$ , $E_p =
\sqrt{M^2 +{\bf P}^2}$  and
$S(\Lambda)$ is the Lorentz transformation matrix for a Dirac spinor
\cite[page 77]{QFT}. We clearly see how we obtain the plane wave
function $e^{-i P \cdot Y}$, which describes the motion of the CM of
the isolated system.  This equation also gives us the Lorentz
transformation rule for the $q(\xi)$'s. Again, there is no internal
time variables $t_a$ in our formula,  which follows from our restriction
on energy eigenstates. The normalization of $q_n(x)$ is Lorentz invariant,


\begin{equation}
 \int d^3x' q'^\dagger_n(x') q'_{n'}(x') = \int \frac{d^3 x'}{\gamma}
q_n^\dagger(x) q_{n'}(x)\gamma = \int d^3x q^\dagger_n(x)q_{n'}(x)  =
\delta_{n,n'}
\label{eq:norm_qn}
\end{equation}
Also we have the  invariant normalization for $q_{nv}(x)$:
\begin{equation}
 \int d^3x q^\dagger_{nv}(x) q_{n'v}(x) =  \delta_{n,n'}
\label{eq:norm_qnv}
\end{equation}
To check, we let $x'=\Lambda x$ and note that $d^3x'=d^3x/\gamma$,
as the measure of a proper volume in moving frame, and $\psi^\dagger
\psi=\bar{\psi}\gamma^0\psi$, which is the zeroth component of a
4-vector. Hence:
\begin{eqnarray*}
\int d^3x'\, q'^\dagger_n(x') q'_{n'}(x')& = &\int \frac{d^3 x}{\gamma}
         q_n^\dagger(x) q_{n'}(x)\gamma = \delta_{n,n'},  \\
\int d^3x \, q^\dagger_{nv}(x) q_{n'v}(x)
     &\equiv & \int d^3x  \, q'^\dagger_{n}( x) q'_{n'}(x) \\
\equiv  \int d^3x \, \gamma q^\dagger_{n}( \Lambda^{-1}x)
q_{n'}(\Lambda^{-1}x)
   &=& \int d^3x' \, \gamma q^\dagger_{n}(x) q_{n'}(x) = \delta_{n,n'}
\end{eqnarray*}
The normalization of $ \psi_{n P}(x_1, \ldots, x_N)$  is similar as in
eq.(\ref{eq:PhiP-P'}):

\begin{equation}
\int d^3 Y d^{3N}\xi \, \psi^\dagger_{n P}\psi_{n' P'} = \delta_{n, n'}
\delta^3({\bf  P} - {\bf P}')
\label{eq: PhiP-P'2}
\end{equation}

Now we want to expand any function $ \Psi(x_1, \ldots, x_N)$ which
represents a moving confined system. Using ${\bf P}=\gamma M{\bf v}$,
we define $|{\bf  v} \rangle = ( \gamma M )^{3/2} |{\bf P} \rangle $,
or $  \phi_P(Y)=(\gamma M)^{-3/2} \phi_v(Y)  $. Thus we have

\[  \langle {\bf P } | {\bf P }' \rangle = \delta^3( {\bf P } - {\bf P }') =
\delta^3( {\bf v } - {\bf v }')/(\gamma M)^3 \]
and we  obtain {\em the revised CM-freedom separation rule}:

\begin{eqnarray}
\Psi(x_1, \ldots, x_N) &=& \int d^3 v \, a_v \sum _{n} \phi_v(Y)
            b_n q_{nv}(\xi_1, \ldots  \xi_N)   \label{eq:sep_v}\\
    & =&  \int d^3 P \, a_P \sum _{n} \phi_P(Y)
            b_n q_{nv}(\xi_1, \ldots  \xi_N)
\label{eq:sep_rule2}
\end{eqnarray}
where we have used  $ a_v = a_P (\gamma M)^{3/2} $ and

\[  d^3v = d^3 P \det(\partial v^i/\partial P^j) = d^3 P/ (\gamma M)^3\].

Note that the expansion in (\ref{eq:sep_v}) is to find all  system states
with ${\bf v} = {\bf 0}$ and then boost each of them with all possible
${\bf v}$, with $ P = (\gamma M, \gamma M{\bf v})$ and   $ P^2 = M^2$.
Thus the expansion is the same either in CMRF or in CMVR. It has the same
advantages as we have mentioned in the non-relativistic study.

\vspace{0.4cm}
{\em 5. The $ \gamma-N$ effective interaction Lagrangian}:
In ref.\cite{CM-0,CM-1,CM-2,CM-3}, using the old separation rule
of eq.(\ref{eq:sep_rule0}),  we have introduced and used an effective
$ \gamma-N$   interaction Lagrangian:

\begin{equation}
L^{\gamma-N}(Y) = J_\mu(Y)A^\mu(Y) = \sum_{1 \rightarrow 2,3} \int
\frac{d^9\xi }{\lambda^3} \Psi(x_1,x_2,x_3)( \hat{e}\gamma_0
\gamma_\mu)_1A^{\mu}(x_1) \Psi(x_1, x_2, x_3)
\label{eq:LY_1}
\end{equation}
 Now  we use our revised one, eq.(\ref{eq:sep_rule2}),
and we find that we do not need to introduce the parameter $\lambda$
to make the action $ S = \int d^4Y \, L^{\gamma-N}(Y)$ dimensionless.
The effective lagrangian now can be written without any undetermined
parameter, namely:

\begin{equation}
L^{\gamma-N}(Y) = J_\mu(Y)A^\mu(Y) = \sum_{1 \rightarrow 2,3} \int
d^9\xi \Psi(x_1,x_2,x_3)( \hat{e}\gamma_0
\gamma_\mu)_1A^{\mu}(x_1) \Psi(x_1, x_2, x_3)
\label{eq:LY_2}
\end{equation}
When eq.(\ref{eq:LY_2}) is applied to calculation of nucleon EM
form-factors \cite{CM-1},  with
 \[ A^\mu(x_1) =  A^\mu (q) e^{ - i q \cdot (Y+\xi_1)} = A^\mu (q)   e^{ - i q
\cdot Y} e^{ i {\bf q \cdot \xi}_1} = A^\mu (Y) e^{ i {\bf q \cdot \xi}_1} \]
and  the normalization $ \langle P' | P \rangle = \delta^3({\bf P' - P})$,
the effective  $ \gamma-N$  vertex is derived as:

\[ \int d^4Y e^{-i q \cdot Y}  \langle P'|J_\mu(Y) | P \rangle
= (2 \pi) \delta^4(P + q - P') \langle P'|J_\mu(0) | P \rangle \]
where:
\begin{equation}
\langle P'|J_\mu(0) | P \rangle  = \sum_{1 \rightarrow 2,3} \int
d^9\xi e^{i {\bf q \cdot \xi}_1} q^\dagger_{v' n'}( {\bf \xi}_1, {\bf \xi}_2,
{\bf \xi}_3)( \hat{e} \gamma_0 \gamma_\mu)_1 q_{vn}({\bf \xi}_1, {\bf \xi}_2,
{\bf
\xi}_3)
\label{eq:J_0P}
\end{equation}
This vertex is exactly the righthand side of eq.(4) or eq.(6) in Ref.[7],
which leads to proton form-factors in quite good agreement with the data.
Hence, we can write our {\em revised effective current} as:

\begin{equation}
J_\mu(Y) =  \sum_{1 \rightarrow 2,3} \int d^9\xi  e^{i {\bf q
\cdot \xi}_1} \Psi(x_1,x_2,x_3)( \hat{e}\gamma_0 \gamma_\mu)_1 \Psi(x_1, x_2,
x_3)
\label{eq:JY}
\end{equation}
which has no undetermined parameter. We note that unlike the ordinary
current expression, where only $one$ volume element $d^3x$ times the
zeroth component of a vector, which together make a Lorentz scalar,
we now have a product of $three$ volume integrals of a zeroth component
of a 4-vector of one struck quark, multiplied by
an invariant (in Breit frame) from other two spectator quarks.
Hence we have an extra factor $1/(ch \Omega)^2=1/(1+Q^2/4M^2)$
left uncanceled. This is the factor which appears in front of the
electric form factor $G_E(Q^2)$ in eq. (11a) in ref.[7].
The current (49) can be easily extended to
more general currents. For example, we can replace the $U(1)$ generator
$\hat{e}$ by $\lambda^a $ of the flavor $SU(3)$ generator
to obtain the $SU(3)$ current $J^a_\mu(Y)$.

\vspace{0.4cm}
{\em 6. The Deep Inelastic $ \gamma-N$ Collision  and Free-Quark
approximation}:  To find nucleon structure functions, we have to begin
with the following tensor:
\begin{eqnarray*}
 W_{\mu\nu} &=& \ \frac{1}{4\pi}\int d^4Y d^3P' \sum_{S'} e^{i q \cdot Y}
\langle P, S | J_\mu(Y)|P', S' \rangle \langle P',S'|J_\nu(0) |P, S \rangle\\
&=&  \frac{1}{4\pi}\int d^4Y e^{i q \cdot y} \langle P, S |[ J_\mu(Y), J_\nu(0)
] |P, S \rangle
\end{eqnarray*}
where $ \langle P, S | P',S' \rangle = (2 \pi)^3 2E \delta^3(P - P') \delta_{S,
S'}$. When using free quark approximation for the ``intermediate" states
above, i.e., assuming  that after the scattering  all quarks go freely,
we will use free quark anticommutators like $ \{\psi(x_i), \bar{\psi}(x_i) \}$
\cite{CM-0,CM-4}. So  we can not use our $J_\mu(Y)$ as defined in
eq.(\ref{eq:JY}), where both incoming and outgoing quarks are in
designative states according to the revised expansion rule of $\Psi $.
We should  do the following replacement in eq.(\ref{eq:JY}):

\begin{eqnarray*}
  \sum_{1 \rightarrow 2,3} \int d^9\xi  e^{i {\bf q
\cdot \xi}_1} \Psi(x_1,x_2,x_3)( \hat{e}\gamma_0 \gamma_\mu)_1 \Psi(x_1, x_2,
x_3)  \\
\rightarrow   \sum_{1 \rightarrow 2,3} \int d^9\xi  e^{i {\bf q
\cdot \xi}_1} \Psi(x_1,x_2,x_3)( \hat{e}\gamma_0 \gamma_\mu)_1
\psi(x_1)\psi(x_2)\psi(x_3)
\end{eqnarray*}
Therefore, according to eq.(\ref{eq:dimension}), we must introduce
a parameter $\lambda$ with dimension $L$, and the effective current
in the free-quark approximation becomes:

\begin{equation}
\tilde{J}_\mu(Y) =  \sum_{1 \rightarrow 2,3} \int d^9\xi  e^{i {\bf q
\cdot \xi}_1} \Psi(x_1,x_2,x_3)( \hat{e}\gamma_0 \gamma_\mu)_1
\frac{1}{\lambda^{3/2}}\psi(x_1)\psi(x_2)\psi(x_3)
\label{eq:JY2}
\end{equation}
In the bag model, the only reasonable choice for $\lambda$ is:

\begin{equation}
\lambda_i = C_i(Q^2) R
\label{eq:lambda}
\end{equation}
Here R  is the bag radius (a Lorentz invariant constant! ), and $ C_i$
is possibly $Q^2$-dependent \cite{CM-0}, because of the factor
$\exp(i{\bf q} \cdot{\bf \xi})$ in $  \tilde{J}$ and it may also
be flavor-dependent \cite{CM-5}. We have compared our result in
the non-relativistic limit with Jaffi's result in ref.\cite{MIT2},
we find  that  $ \lambda = R $ \cite{CM-01}.  In ref. \cite{CM-5},
$\lambda$'s for a proton are fixed  through the rms radius of the
neutron and proton,  with the results: $\lambda_u = R $ and
$\lambda_d = 0.85R $.

In ref.\cite{CM-4}, some Feynman rules for the CM-bag model are
given. We see that  parameter $\lambda $ does not appear in the
$\gamma-N$ vertex anyway. Only in the last two graphs, where we
have used free-quarks for out-going states, $\lambda $ is not
canceled out. So our Feynman rules remain unchanged. In the same
paper, we also mentioned that the bag radius $R$ might be
$Q^2$-dependent to explain the EMC effect.

   For other models, we can introduce similar currents. When we
use free particle approximation for out-going particles, we need
to introduce $\lambda$ as in eq.({\ref{eq:JY2}), with $\lambda$
proportional to the length scale given by the model.  For example,
if the potential is an isotropic harmonic oscillator potential
$V(r)= k r^2/2 = m \Omega^2 r^2/2 $, then in our semi-relativistic
approach,  we have only one  length scale (through three constants
$m, \Omega, \hbar=1)$,  namely

\begin{equation}
 \lambda_i =\frac{ C(Q^2)} {\sqrt{m_i\Omega}}
\label{eq:Losci}
\end{equation}
For a relativistic oscillator \cite{ROsci1,ROsci2},  the way to
get a length scale is not unique ( note, e.g.,  that $[ c \Omega ]
=[ \sqrt{\hbar /(m\Omega)} ] =L$ ), but if we look at the parameter
$\lambda_{Nj\pm}$ used to define the dimensionless coordinate
$ r' = \lambda  r$ in ref.\cite{ROsci2}, we find eq.(\ref{eq:Losci})
is still true.

\vspace{0.4cm}
{\it 5. Summary and Discussion}:  We find that if we assume a
central field, and restrict to designative states, our CM 4-vector
is well-defined. We have revised our CM-freedom separation rule
and the effective $\gamma-N$ current, neither of them now have
undetermined parameters. The length scale $\lambda$ comes in only
when we use the free-particle approximation for deep inelastic
scatterings.

Our method can be applied to the CM bag model and any other model
with central field.  A very interesting case will be  an isotropic
harmonic oscillator potential. This can be used for a (non-)
relativistic nuclear shell model or an alternative hadron model
with quarks(antiquarks) of  non-zero rest mass (so the $SU(3)$
flavor asymmetry can be easily introduced). We will discuss the
nucleon structure functions given by such a hadron model in our
future work.

\pagebreak

\end{document}